\documentclass[aps,twocolumn,showpacs,preprintnumbers,superscriptaddress,prb]{revtex4}

\usepackage{graphicx}
\usepackage{dcolumn}
\usepackage{bm}

\begin{document}


\title{Metal Surface Energy: Persistent Cancellation of Short-Range\\
Correlation Effects beyond the Random-Phase Approximation}  

\author{J. M. Pitarke}
\affiliation{Materia Kondentsatuaren Fisika Saila, Zientzi Fakultatea,
Euskal Herriko Unibertsitatea, 644 Posta Kutxatila, E-48080 Bilbo, Basque
Country}
\affiliation{Donostia International Physics Center (DIPC) and Centro Mixto
CSIC-UPV/EHU, Donostia, Basque Country}                                         
\author{J. P. Perdew}
\affiliation{Department of Physics and Quantum Theory Group, Tulane
University, New Orleans, LA 70118}

\date{\today}

\begin{abstract}
The role that non-local short-range correlation plays at metal surfaces is
investigated by analyzing the correlation surface energy into contributions
from dynamical density fluctuations of various two-dimensional wave vectors.
Although short-range correlation is known to yield considerable
correction to the ground-state energy of both uniform and non-uniform
systems, short-range correlation effects on intermediate and short-wavelength
contributions to the surface formation energy are found to compensate one
another. As a result, our calculated surface energies, which are based on a
non-local exchange-correlation kernel that provides accurate total energies of
a uniform electron gas, are found to be very close to those obtained in the
random-phase approximation and support the conclusion that the error
introduced by the local-density approximation is small.    

\end{abstract}


\maketitle

\section{Introduction}

The widely-used Kohn-Sham formulation of density-functional theory (DFT)
\cite{ks} requires approximations to the exchange-correlation (xc) energy
$E^{xc}[n({\bf r})]$.  The simplest approximation to this functional is the
so-called local-density approximation (LDA), where $E^{xc}[n({\bf r})]$ is
given at each point by the xc energy of a uniform electron gas at the local
density. This approximation was found to be remarkably accurate in some rather
inhomogeneous situations, \cite{gunnarsson} and its widespread use in
condensed-matter physics led to the early success of DFT.

Hence, it is important that the LDA be tested against {\it benchmark} systems,
such as the jellium surface, and that new functionals be developed.
Nevertheless, more than 30 years after Lang and Kohn reported the first
self-consistent LDA calculation of the jellium surface energy, \cite{lk} the
question of the impact of non-local xc effects on the surface energy and their
interplay with the strong charge inhomogeneity at the surface has remained a
puzzle. \cite{maziar,new0} The simple LDA and more advanced density functionals
such as generalized gradient approximations (GGA's) \cite{gga} and meta-GGA's
\cite{mgga} all predict the same jellium xc surface energy within a few
percent, but show no such agreement with the available
wave-function based methods: Fermi hypernetted chain (FHNC) \cite{kk} and
diffusion Monte Carlo (DMC); \cite{acioli2} see Table I of Ref.
\onlinecite{zidan}.

An alternative formally exact way to find the xc energy of an arbitrary
inhomogeneous system is provided by the adiabatic connection
formula and the fluctuation-dissipation theorem. \cite{langreth}
Within this approach, the exchange energy is fully determined from the
exact Kohn-Sham (KS) orbitals and the correlation energy is obtained in terms
of the xc kernel $f^{xc}$. \cite{gk} In the random-phase approximation (RPA),
$f^{xc}$ is taken to be zero. Full RPA or corrected-RPA calculations are now
feasible not only for bulk jellium \cite{lein} but also for jellium
surfaces \cite{pitarke} and molecules. \cite{f,fg,amt}

In this paper, we take a non-local xc kernel that provides accurate
ground-state energies of a uniform electron gas, \cite{lein} and evaluate the
jellium surface energy through the use of the adiabatic connection formula. We
analyze the correlation surface energy into contributions from dynamic density
fluctuations of various two-dimensional wave vectors, and find that
short-range xc effects on intermediate and short-wavelength fluctuations
nearly compensate. Hence, while RPA is known to be a poor approximation for
the total correlation energy, our calculations show that it is a surprisingly
good approximation for those changes in the correlation energy that arise in
surface formation. This is in contrast with FHNC and DMC slab calculations,
\cite{kk,acioli2} which predict surface energies that are significantly
higher than those obtained either in the LDA \cite{lk} or in a fully non-local
RPA. \cite{pitarke}

The FHNC variational equations have been shown to provide, in the homogeneous
limit, reasonable agreement with the known properties of a uniform electron
gas, \cite{krot} and DMC calculations are often regarded as essentially exact.
\cite{rmp} Furthermore, one may expect that when applied to non-uniform
systems these wave-function-based approaches will lead to results whose
accuracy is comparable to the high accuracy obtained for uniform systems.
Nevertheless, we show that surface formation energies obtained from slab
calculations either by a linear fit in the slab thickness \cite{kk} or as
differences between slab energies and an independently determined bulk energy
\cite{acioli2} may result in substantial imprecision, and conclude
that wave-function-based estimates need to be reconsidered.

\section{Theoretical framework}

We consider a jellium slab normal to the $z$ axis, which is translationally
invariant in the surface plane. Subtracting from the slab energy
the corresponding energy of a uniform electron gas and using the adiabatic
connection formula, one obtains the xc surface energy
\begin{equation}\label{surf1}
\sigma^{xc}=\int_0^\infty d(q/\bar k_F)\,\gamma_q^{xc},
\end{equation}
${\bf q}$ being a wave vector parallel to the surface. $\bar
k_F$ is the Fermi momentum and $\gamma_q^{xc}$ represents the surface energy
associated with the good quantum number $q$: 
\begin{eqnarray}\label{surf2}
\gamma_q^{xc}&=&\frac{\bar k_F}{8\,\pi}\int dz\int
dz'\,n(z)\,v_q(|z-z'|)\nonumber\\
&\times&\int_0^1 d\lambda\left[n_{q,\lambda}^{xc}(z,z')- \bar
n_{q,\lambda}^{xc}(|z-z'|)\right] , 
\end{eqnarray}
$v_q(|z-z'|)=(2\pi e^2/q)\exp\left(-q|z-z'|\right)$ being the Fourier transform
of the bare Coulomb interaction. $n_{q,\lambda}^{xc}(z,z')$ and $\bar
n_{q,\lambda}^{xc}(|z-z'|)$ represent Fourier components of the xc-hole density
of a fictitious jellium slab at coupling strength $\lambda e^2$ and the
corresponding xc-hole density of a uniform electron gas of density $\bar
n=\bar k_F^3/3\pi^2$, respectively. In the LDA, the xc surface energy is
obtained by simply replacing $n_{q,\lambda}^{xc}(z,z')$ by the xc-hole density
of a uniform electron gas of density $n(z)$. A parametrization of the
uniform-gas xc-hole density has been given by Perdew and Wang, \cite{pw2}
which  yields the DMC ground-state energy of a uniform electron gas. \cite{ca}

According to the fluctuation-dissipation theorem, 
\begin{equation}\label{fd}
n_{q,\lambda}^{xc}(z,z')=-\frac{\hbar}{\pi\,n(z)}\int_0^\infty
d\omega\,\chi_{q,\lambda}(z,z';i\,\omega)-\delta(z-z'),
\end{equation}
where $\chi_{q,\lambda}(z,z';\omega)$ is the interacting density-response
function. Time-dependent DFT (TDDFT) shows
that this function obeys the Dyson-type equation \cite{tddft,dobson}
\begin{eqnarray}
&&\chi_{q,\lambda}(z,z';\omega)=\chi_q^0(z,z';\omega)+\int dz_1\int
dz_2\, \chi_q^0(z,z_1;\omega)\nonumber\\
&\times&\left[\lambda\,v_q(|z_1-z_2|)+f_{q,\lambda}^{xc}[n](z_1,z_2;\omega)\right]
\chi_{q,\lambda}(z_2,z';\omega), \end{eqnarray}
where $\chi_q^0(z,z';\omega)$ is the density-response function of
non-interacting KS electrons \cite{note} and
$f_{q,\lambda}^{xc}[n](z,z';\omega)$ involves the functional derivative of the
KS xc potential at coupling constant $\lambda$.

Using the coordinate-scaling relation for the $\lambda$-dependence of the xc
kernel derived in Ref. \onlinecite{lein}, we find \begin{equation}
f_{q,\lambda}^{xc}[n(z)](z,z';\omega)=f_{q/\lambda}^{xc}[\lambda^{-3}\,n(z/\lambda)] (\lambda z,\lambda z';\omega/\lambda^2), 
\end{equation} 
where $f_q^{xc}[n](z,z';\omega)$ is the xc kernel at $\lambda=1$. In order
to derive an approximation for this quantity, we assume that the density
variation [$n(z)-n(z')$] is small within the short range of
$f_q^{xc}[n](z,z';\omega)$ and write \cite{lein}
\begin{equation}\label{fxc}
f_q^{xc}[n](z,z';\omega)=\bar
f_q^{xc}(\left[n(z)+n(z')\right]/2;|z-z'|;\omega), \end{equation}
where $\bar f_q^{xc}(n;|z-z'|;\omega)$ is the Fourier transform of
the xc kernel $\bar f^{xc}(n;k,\omega)$ of a uniform electron gas of density
$n$. Here ${\bf k}=({\bf q},k_z)$ represents a three-dimensional wave vector.

\section{Results and discussion}

We have carried out simplified surface-energy calculations with $\bar
f_q^{xc}(n;|z-z'|;\omega)$ replaced by $\bar f^{xc}(n;k=q,\omega)\delta(z-z')$
[thus assuming that the dynamic density fluctuation is slowly varying in the
direction perpendicular to the surface] and using the parametrization of
Richardson and Ashcroft for $\bar f^{xc}(n;k,\omega)$, \cite{ra} and have
found that neglect of the frequency dependence of the xc kernel does not
introduce significant errors. We have also carried out adiabatic LDA (ALDA)
surface-energy calculations with   $\bar f_q^{xc}(n;|z-z'|;\omega)$ replaced by
$\bar f^{xc}(n;k=0,\omega=0)\delta(z-z')$ [thus assuming that the dynamic
density fluctuation is slowly varying in all directions], and have found that
the spacial range of the xc kernel cannot be neglected. These conclusions also
apply to the uniform electron gas. \cite{lein}

Hence, we neglect the frequency dependence of the xc kernel and exploit the
accurate DMC calculations reported in Ref. \onlinecite{moroni} for the static
xc kernel of a uniform electron gas. A parametrization of this
data satisfying the known small- and large-wavelength asymptotic behavior has
been reported, \cite{corradini} which allows us to write \begin{widetext}
\begin{equation}\label{corradini}
\bar f_q^{xc}(n;|z-z'|)=-\frac{4\pi e^2 C}{k_F^2}\delta(\tilde z)-\frac{2\pi
e^2 B}{\sqrt{gk_F^2+q^2}}\,{\rm e}^{-\sqrt{gk_F^2+q^2}|\tilde z|}-
\frac{2\alpha\sqrt{\pi/\beta}e^2}{k_F^3}\left[\frac{2\beta-k_F^2\tilde
z^2}{4\beta^2}k_F^2+q^2\right] {\rm e}^{-\beta\left[k_F^2\tilde z^2/4\beta^2
+q^2/k_F^2\right]},
\end{equation}
\end{widetext}
where $C$, $B$, $g$, $\alpha$, and $\beta$ are dimensionless functions of the
electron density (see Ref. \onlinecite{corradini}), $n=k_F^3/3\pi^2$,
and $\tilde z=z-z'$. The finite $q\to 0$ limit of Eq. (\ref{corradini}) will be
dominated by the $q\to 0$ divergence of $v_q(|z-z'|)$, making RPA exact in
this limit. In the large-$q$ limit, where short-wavelength excitations tend
to be insensitive to the electron-density inhomogeneity, introduction of Eq.
(\ref{corradini}) into Eq. (\ref{fxc}) is expected to yield an xc kernel that
is essentially exact.

If the interacting density-response function $\chi_{q,\lambda}(z,z';\omega)$
entering Eq. (\ref{fd}) is replaced by  $\chi_{q}^0(z,z';\omega)$, Eqs.
(\ref{surf1}) and (\ref{surf2}) yield the {\it exact} exchange surface energy,
as obtained in Ref. \onlinecite{pitarke}. Here we focus our attention on
the correlation surface energy, which for comparison we also calculate in
LDA by replacing $n_{q,\lambda}^c$ in Eq. (\ref{surf2}) by the uniform-gas
correlation-hole density at the {\it local} density $n(z)$.

\begin{figure}
\includegraphics[width=0.45\textwidth,height=0.3375\textwidth]{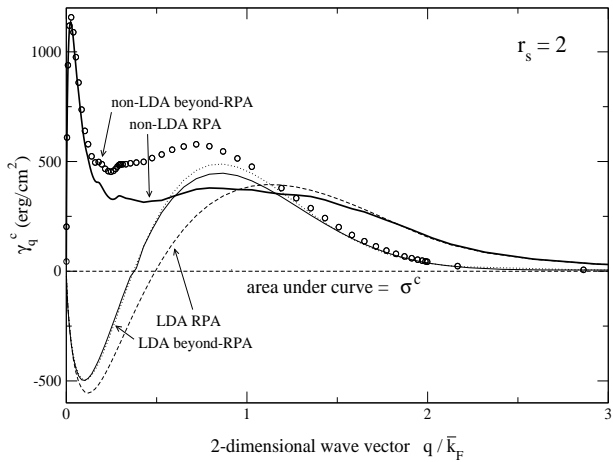}
\caption{Wave-vector analysis $\gamma_q^{c}$ of the correlation surface
energy for a jellium slab of thickness $a=7.17r_s$ and $r_s=2$. Thin-solid,
dashed, and dotted lines represent Perdew-Wang, RPA and Corradini-based LDA
calculations, respectively. The thick-solid line and the open circles
represent RPA and Corradini-based non-LDA calculations, respectively.  Our
'best' non-local calculation (open circles) provides the {\it exact} small-$q$
and large-$q$ limits.} \label{fig1}
\end{figure} 

Fig. 1 shows the wave-vector analysis $\gamma_q^c$ of both LDA and non-LDA
correlation surface energies of a jellium slab of thickness $a=7.17r_s$ and
$r_s=2$. \cite{note3} First of all, we focus on our LDA calculations, which
have been carried out either by using the uniform-gas correlation-hole density
with $\bar f_q^{xc}(n;|z-z'|)=0$ (RPA-based LDA) or the xc kernel of Eq.
(\ref{corradini}) (Corradini-based LDA), or by using the Perdew-Wang (PW)
parametrization of Ref. \onlinecite{pw2}. We observe that in the
long-wavelength limit ($q\to 0$) both RPA and Corradini calculations coincide
with the PW parametrization. At shorter wavelengths, the Corradini scheme
predicts a substantial correction to the RPA and accurately reproduces the
PW wave-vector analysis of the correlation energy. Hence, armed with some
confidence in the accuracy of our choice of the xc kernel, we apply it to the
more realistic non-local scheme described above. Fig. 1 shows that our non-LDA
beyond-RPA ('best' non-local) calculation coincides in the $q\to 0$ limit
\cite{note4} with the non-LDA RPA, which is exact in this limit. In the
large-$q$ limit, local and non-local calculations coincide, and
our 'best' non-local calculation accurately reproduces the PW-based LDA
(thin solid line), which is expected to be esentially exact in this limit.

Consequently, our 'best' non-local calculation provides both the {\it exact}
small-$q$ limit, where LDA fails badly, and the {\it exact} large-$q$ limit,
where RPA is wrong. The LDA  largely underestimates our non-local correlation
surface energy, but Fig. 1 shows that the difference between RPA and
beyond-RPA $\gamma_q^c$ (the short-range part of the correlation surface
energy) is fairly insensitive to whether the LDA is used or not. This
supports the assumption made in Ref. \onlinecite{kurth2} that the short-range
part of the correlation energy can be treated within LDA or GGA. However,
short-range xc effects on intermediate and short-wavelength contributions to
the surface energy tend to compensate, and this cancellation happens to be even
more complete than expected from LDA or GGA. Thus, our non-local scheme yields
surface energies which are still closer to RPA than is the RPA$^+$ of Ref.
\onlinecite{kurth2}.

\begin{table}
\caption{Non-local xc ($\sigma^{xc}$) and total ($\sigma$) surface energies
and their local (LDA) counterparts, as obtained from Eqs. (\ref{surf1}) and
(\ref{surf2}) with the non-local xc kernel of Eq. (\ref{corradini}). Also
shown are non-local RPA xc surface energies ($\sigma_{\rm RPA}^{xc}$) and
PW-LDA total surface energies ($\sigma_{\rm PW-LDA}$). Tiny differences
between these RPA xc surface energies and those reported before \cite{pitarke}
are entirely due to differences in the parametrization of the xc potential.
Units are erg/cm$^2$.}  \begin{ruledtabular} \begin{tabular}{lcccccc}
$r_s$&$\sigma^{xc}_{\rm RPA}$&$\sigma^{xc}$ &$\sigma^{xc}_{\rm
LDA}$&$\sigma$&$\sigma_{\rm LDA}$&$\sigma_{\rm PW-LDA}$\\ \hline
2.00&3467&3466&3369&-752&-849&-862\\ 2.07&3064&3063&2975&-504&-592&-605\\
2.30&2098&2096&2026&-27&-97&-103\\
2.66&1242&1239&1193&221&175&171\\
3.00&803&797&767&258&228&225\\
3.28&580&577&551&247&221&220\\
4.00&279&277&262&179&164&164\\
5.00&120&119&111&106&98&98\\
6.00&59&58&54&64&59&59\\
\end{tabular}
\end{ruledtabular} \label{table1}
\end{table}

To extract the surface energy of a semi-infinite medium, we have considered
three different values of the slab thickness: the threshold width at which
the $n=5$ subband for the $z$ motion is completely occupied and the two widths
at which the $n=5$ and $n=6$ subbands are half occupied, and have followed
the extrapolation procedure of Ref. \onlinecite{pitarke}. In Table I we show
our extrapolated local (LDA) and non-local surface energies, as obtained from
Eqs. (\ref{surf1}) and (\ref{surf2}) either with $\bar f_q^{xc}(n;|z-z'|)=0$
(RPA) or with the xc kernel of Eq. (\ref{corradini}). These
calculations indicate that the introduction of a {\it plausible} non-local xc
kernel yields short-range corrections to RPA surface energies that are
negligible. For comparison, also shown in Table I are PW-LDA surface energies,
as obtained either from Eqs. (\ref{surf1}) and (\ref{surf2}) with the
PW uniform-gas xc-hole density \cite{pw2} or from the PW
parametrization of the uniform-gas xc energy. \cite{pw1} Corradini and
PW-based LDA surface energies ($\sigma_{\rm LDA}$ and $\sigma_{\rm PW-LDA}$)
are found to be very close to each other, and we expect our Corradini-based
non-local surface energies ($\sigma$) to be close to the {\it exact} jellium
surface energy, as well.

We close this paper with an analysis of the available wave-function-based
surface-energy calculations. Krotscheck {\it et al.} \cite{kk2} considered
slabs of four different thickness $a$, and obtained both a bulk energy per
particle $\varepsilon_\infty$ and a surface energy $\sigma$ from the FHNC slab
energy per particle $\varepsilon$ as a function of the particle number per
unit area $\bar na$, by a linear fit $\varepsilon(\bar
na)=\varepsilon_\infty+2\sigma/\bar na$ that becomes exact in the limit of
infinite thickness. These authors showed that the extrapolated bulk energies
($\varepsilon_\infty$) and those obtained from a separate FHNC bulk
calculation ($\bar\varepsilon$) agree within about $1\%$, and claimed that
this comparison lent credibility to their numerical treatment. However, these
small differences in the bulk calculation yield an uncertainty in the surface
energy $\Delta\sigma=(\bar\varepsilon-\varepsilon_\infty)\bar na/2$, which for
$r_s=2.07$ and 4.96 can be as large as 280 and 11 erg/cm$^2$, respectively.
Moreover, since for the smallest/largest width under study and due to
oscillatory quantum-size effects the quantity $\varepsilon(\bar na)$ is
larger/smaller than expected for a semi-infinite medium, the extrapolated bulk
and surface energies are found to be too negative and too large, respectively.

Li {\it et al.} \cite{acioli1} calculated the fixed-node DMC surface energy of
a jellium slab with $r_s=2.07$ and found $\sigma=-465\,{\rm erg/cm^2}$, which
is $\sim 40\,{\rm erg/cm^2}$ larger than the RPA value. They also performed LDA
calculations with either the Wigner or the Ceperley-Alder form for the
uniform-gas xc energy, and found LDA surface energies that are also about
$40\,{\rm erg/cm^2}$ larger than the corresponding LDA surface energies of a
semi-infinite jellium, which suggests that finite-size corrections might bring
the DMC surface energy into close agreement with RPA. These fixed-node DMC
calculations were extended by Acioli and Ceperley to study jellium slabs at
five different densities, \cite{acioli2} but these authors extracted the
surface energy from release-node bulk energies. Both Li
{\it et al.} \cite{acioli1} and Acioli and Ceperley \cite{acioli2} claimed
that the release-node correction of the uniform electron gas at $r_s=2.07$ is
$0.0023\,{\rm eV/electron}$, and argued that this correction would only yield
a small error in the surface energy. However, the unpublished uniform-gas
fixed-node energy reported and used in Ref. \onlinecite{acioli1} is actually
$0.0123\,{\rm eV/electron}$ higher than its release-node counterpart; hence,
by combining fixed-node slab and release-node bulk energies Acioli and
Ceperley produced for $r_s=2.07$ a surface energy that is too large by 138
erg/cm$^2$. Furthermore, had these authors used fixed-node bulk energies (see,
e.g., Ref. \onlinecite{martin}), they would have obtained surface energies that
are close to RPA.

\section{Summary and conclusions}

We have investigated the role that short-range correlation
plays at metal surfaces, on the basis of a wave-vector analysis of the
correlation surface energy. Our non-local calculations, which are found to
provide the {\it exact} small and large-$q$ limits, indicate that a
persistent cancellation of short-range correlation effects yields surface
energies that are in excellent agreement with RPA, and support the conclusion
that the error introduced by the LDA is small. Although this conclusion seems
to be in contrast with available wave-function based calculations, we have
shown that a careful analysis of these data might bring them into close
agreement with RPA. This is consistent with recent work, where jellium surface
energies extracted from DMC calculations for jellium spheres are also found to
be close to RPA. \cite{new}

We have found that the RPA xc surface energy displays an error
cancellation between short- and intermediate-range correlations. A different
(and less complete) error cancellation between long- and intermediate-range xc
effects explains \cite{zidan} why the LDA works for the surface energy. The GGA
corrects only the intermediate-range contributions, \cite{zidan} and so gives
surface energies slightly lower and less accurate than those of LDA. Usually
GGA works better than LDA, but not for the jellium surface energy where
long-range effects are especially important. In the present work, as in four
others, \cite{zidan,kurth2,new,kpb} we have found that the jellium xc surface
energy is only a few percent higher than it is in LDA. These closely-agreeing
methods include two different short-range corrections to RPA (present work
using a non-local xc kernel and Ref. \onlinecite{kurth2} using an additive
GGA correction), a long-range correction to GGA (Ref. \onlinecite{zidan}),
extraction of a surface energy from DMC energies for jellium spheres (Ref.
\onlinecite{new}), and a meta-GGA density functional (Ref. \onlinecite{mgga}).
The corresponding correction to the LDA or GGA surface energy has been
transferred \cite{mk} successfully to the prediction of vacancy formation
energies \cite{carling} and works of adhesion. \cite{mat}

We have almost reached a solution of the surface-energy puzzle, but one piece
still does not fit. Krotscheck and Kohn \cite{kk2} examined a "collective RPA"
which approximates our full RPA, and also used several xc kernels to correct
for short-range effects. When they used an isotropic xc kernel derived from the
uniform gas, in the spirit of our Eq. (6), they found surface energies very
close to RPA, as we do. When they used FHNC,\cite{kk} corresponding to an
anisotropic (but to the eye not very different) kernel constructed explicitly
for the jellium surface, they found a large positive correction to the RPA
surface energy, amounting at $r_s=4$ to as much as $35\%$ of the RPA
$\sigma_{xc}$ or $60\%$ of the RPA total $\sigma$. The story of the jellium
surface energy cannot reach an end until their work is reconciled with the
other work on this subject. The DMC energies of jellium slabs should also be
re-considered. \cite{foulkes}

Measured surface energies of real metals have been compared with calculations
in Refs. \onlinecite{vitos,ak,fp}. However, experimental and
calculational uncertainties and differences between jellium and real-metal
surfaces seem to preclude a solution to the surface-energy puzzle from these
comparisons.

\acknowledgements   

We wish to acknowledge partial support by the University
of the Basque Country, the Spanish Ministerio de Educaci\'on y Cultura, and
the U.S. National Science Foundation (Grants Nos. DMR-9810620 and DMR-0135678).


\begin{thebibliography}{100}
\bibitem{note} Although $\chi^0$ should be determined from the exact
KS eigenfunctions and eigenvalues, the error introduced by our use of LDA
orbitals is expected to be negligible. Wherever the uniform-gas xc energy is
required as input, we use the parametrization by Perdew and Wang \cite{pw1} of
the Ceperley-Alder DMC data. \cite{ca}
\bibitem{note3} The electron-density parameter
$r_s$ is defined by the relation $1/\bar n=(4/3)\pi r_s^3$.
\end{thebibliography}
\end{document}